\documentclass[aps,prb,twocolumn,amsmath,amssymb,nofootinbib,eqsecnum,tighten,showpacs,floatfix]{revtex4}
\usepackage[dvips]{color} 
\usepackage{graphicx}% Include figure files
\usepackage{dcolumn} % Align table columns on decimal point
\usepackage{bm}      % bold math
\usepackage{epic}
\usepackage{longtable}
\begin{document}

%\preprint{APS/123-QED}

\title{Correlation effects of carbon nanotubes at boundaries:\\
Spin polarization induced by zero-energy boundary states
}

% Force line breaks with \\

\author{ Shinsei Ryu and Yasuhiro Hatsugai}
%\email{Christopher.Mudry@psi.ch}
%\homepage{http://people.web.psi.ch/mudry}
\affiliation{
Department of Applied Physics, University of Tokyo, 7-3-1 Hongo Bunkyo-ku, 
Tokyo 113-8656, Japan 
}
%\altaffiliation[Also at ]{$^2$PRESTO, JST, Saitama 332-0012, Japan}
%\affiliation{%

%\affiliation{

%\date{January 31 2002}
\date{\today}

\begin{abstract}
When a carbon nanotube is truncated 
with a certain type of edges,
boundary states localized near the edges appear at the Fermi level.
Starting from lattice models,
low-energy effective theories 
are constructed 
which describe electron correlation effects on the boundary states.
We then focus on a thin metallic carbon nanotube which
supports one or two boundary states 
and discuss physical consequences 
of the interaction between the boundary states and bulk collective excitations.
By the renormalization group analyses 
together with the open boundary bosonization,
we show that the repulsive bulk interactions
suppress the charge fluctuations at boundaries
and assist the spin polarization.

\end{abstract}

%73.20.At Surface states, band structure, electron density of states
%73.43.Nq Quantum phase transitions
%74.20.-z Theories and models of superconducting state
%73.43.Cd Theory and modeling
%72.80.Rj Fullerenes and related materials
%Use showkeys class option if keyword
\maketitle

%\draft
%\preprint{Dec 07, 2001}

\section{Introduction}

A single-wall carbon nanotube (CNT) is a fascinating
quasi-one-dimensional (quasi-1D) nano-scale material,
which is a graphite sheet wrapped into a cylindrical form.
Its electronic structure is basically well described by 
a one-electron tight-binding model
with a single $\pi$ orbital per atom.
Depending on its geometrical shape,
a large variety of electronic structures 
are realized.
Especially, it can be either 
a metal or a semiconductor
depending on how the graphite sheet is wrapped.
\cite{hamada92,saito92,mintmire92}
The way of wrapping is specified by a chiral vector $(N,M),N,M\in \mathbb{Z}$.
When $N-M \equiv 0 \mod 3$ is satisfied,
a CNT is metallic,
while it is gapped otherwise.

An interesting consequence of its rich electronic band structure
is the existence of the boundary states when the system possesses 
boundaries.
For a CNT with zigzag or bearded edges,
there appear states localized at the boundaries 
for specific values of the wave number along the boundaries.
\cite{fujita96}
It is a hallmark of the phase degree of freedom specific
to quantum mechanical systems.
\cite{ryu02}
The existence of such boundary states 
raises an interesting question as to 
what kind of physical consequences they lead.
For example,
the electronic and magnetic properties of nanographite in magnetic field
\cite{wakabayashi99}
or 
electronic transport through nanographite ribbon junctions
%zero-conductance resonances in nanographite ribbon junctions
\cite{wakabayashi00,wakabayashi01}
were theoretically investigated.
Furthermore,
in the presence of electron-electron or electron-phonon interactions,
the boundary states might trigger an instability
as they form a flatband 
and a sharp peak in density of states at the Fermi energy
for a 2D sheet geometry.
Indeed,
spin polarization induced by the boundary states
\cite{fujita96,wakabayashi98,okada01,hikihara02,nakada98},
or coupling with lattice distortions
\cite{fujita97,ryu02}
has been studied for a graphite sheet 
by several authors.
The effects of 1D low-lying excitations localized at the boundaries
was also discussed 
for ribbon geometry.
\cite{lee02}

The effects of bulk electron correlations
have been extensively investigated 
for metallic CNT's without boundaries.
\cite{egger97,egger98,yoshioka99,odintsov99,krotov97,balents97,lin98,kane97}
It is claimed that 
the most important forward scattering part of 
the Coulomb interaction is well accounted for by 
the Tomonaga-Luttinger (TL) liquid picture
\cite{kane97},
where low-lying excitations are not of Fermi liquid type,
but bosonic collective excitations.
Behaviors specific to TL liquid,
such as a characteristic temperature dependence of
conductance,
have been indeed observed in 
recent experiments.
\cite{bockrath99}
In TL liquid,
boundary critical phenomena 
are known to be drastically different from 
the conventional Fermi liquid case
when the system possesses boundaries.
\cite{kane92,egger92,furusaki93,wang94,fabrizio95}
Then, we expect to see
interesting phenomena
for a thin metallic CNT with boundaries.
The anomalous boundary physics 
in a metallic CNT within the TL liquid picture
such as
tunneling density of states 
\cite{egger98},
Freidel oscillation 
\cite{yoshioka02},
or
local density of states
\cite{yoshioka02-2}
has been investigated previously,
but without boundary states.

The purpose of the present paper is
to discuss electron correlation effects 
for CNT's with edges that supports boundary states.
We consider $(N, -N)$ CNT's 
with zigzag and bearded edges, for which
boundary states appear at the Fermi level
for some values of the wave number along the edges.
Starting from lattice models
with the Coulomb or the Hubbard interaction,
we first establish low-energy effective theories
that describe correlation effects at boundaries.
We then focus on a thin metallic CNT,
where 
the boundary states interact with the collective bulk excitations.
By the renormalization group (RG) analyses together with
the open boundary bosonization,
we discuss the cases where one 
or two boundary states exist.
%paying attentions to roles played by bulk electron interactions.

For the case of two boundary states, we are especially interested in 
whether or not 
the boundary states exhibit spin polarization
in the presence of gapless bulk excitations.
The possibility of spin polarization was
previously discussed for 2D geometry,
i.e., in the limit of the infinite tube radius
$N\to +\infty$,
by mean field theory\cite{fujita96}
or density-functional theory
with local spin density approximation
(LSDA)
\cite{okada01}.
However, these treatments of electron correlations 
can overestimate the magnetic instability.
Also, 
when the system is wrapped into a 1D cylinder,
one needs further justification.
A density matrix renormalization (DMRG)
study was also performed \cite{hikihara02}
which reports spin polarization at boundaries,
but only for a semi-conducting CNT.
When the bulk part of the system is gapless,
it is not clear if the boundary states
show spin polarization
since the total spin (and charge)
carried by the boundary states
can dissipate into a bulk part of the system
through the electron interactions
between boundary states and bulk gapless collective modes.

The paper is organized as follows:
section \ref{Zero energy boundary states and interactions}
is devoted to a construction of low-energy effective theories
which account for zero-energy boundary states.
We recall some known facts on boundary states
in section \ref{Zero energy boundary states},
and the Coulomb or the Hubbard interaction is projected 
to the low-energy sector of the Hilbert space spanned by 
the boundary states and gapless bulk modes
in section \ref{Interactions}.
In section \ref{Effects of the bulk interactions:
Open boundary bosonization},
we perform open boundary bosonization 
to account for the bulk repulsive interactions
and 
discuss its consequences on boundary physics.
Finally, we present RG analyses for 
the case with one boundary states 
(section \ref{Case of one boundary state})
and 
with two boundary states
(section \ref{Case of two boundary state}).
We conclude in section \ref{Conclusion}.

\begin{figure}
\begin{center}
\includegraphics[width=8.5cm,clip]{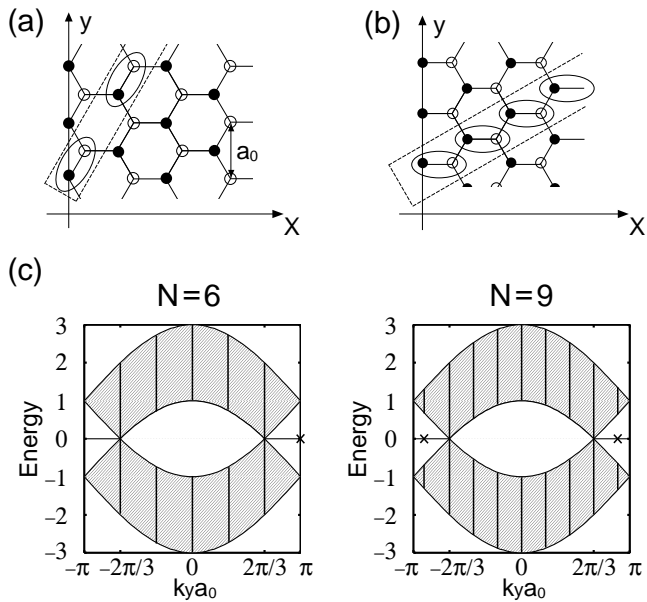}
\caption{
\label{cnt}
A $(N,-N)$ CNT with (a) zigzag and (b) bearded edges.
Two-dimensional graphite sheets are wrapped around 
the $x$ axis.
The ovals 
indicate how to form a spinor $\boldsymbol{c}=\ ^{t}(c_{\bullet},c_{\circ})$,
and dotted squares show a choice of unit cell
for each types of edges.
(c) Energy spectra for a $(N,-N)$ CNT with zigzag edges for $N=6,9$.
($t=1$)
Shaded regions represent bulk energy spectra
for $N\rightarrow \infty$.
Allowed wave numbers along the boundaries for $N=6,9$ are shown
by vertical lines.
Boundary states are denoted by $\times$.
}
\end{center}
\end{figure}

\section{Zero energy boundary states and interactions}
\label{Zero energy boundary states and interactions}

\subsection{Zero energy boundary states}
\label{Zero energy boundary states}

\subsubsection{Zigzag edges}

Let us first consider a $(N,-N)$ CNT with zigzag
edges. 
Our starting point is the single particle
tight-binding Hamiltonian defined on the honeycomb lattice
[Fig. \ref{cnt}(a)]:
\begin{eqnarray}
H_{\hbox{\tiny{kin}}}
&=&\sum_{i,j}
\Big[
\boldsymbol{c}^{\dagger}(i,j)\,
T_{-}^{}\,
\boldsymbol{c}^{}(i,j+1)
+
h.c.
\nonumber \\
&&
+
\boldsymbol{c}^{\dagger}(i,j)\,
T_{+}^{}\,
\boldsymbol{c}^{}(i-1,j+1)
+h.c.
\nonumber \\
&&
+
\boldsymbol{c}^{\dagger}(i,j)\,
T_{0}^{}\,
\boldsymbol{c}^{}(i,j)
\Big],
\end{eqnarray}
where
$\boldsymbol{c}=\ ^{t}(c_{\bullet},c_{\circ})$,
is a spinor made of 
electron annihilation operators $c_{\bullet,\circ}$
defined on different sublattices $\bullet,\circ$.
Coordinates of the spinors are labeled by
the unit cell to which they belong and
the location within the unit cell.
Unit cells are chosen so as to be compatible with
the shape of the edges,
and are labeled by their $y$ coordinate 
$y= j a_{0}$,
$j=1,\cdots,N_{y}$,
where $N_{y}=N$ is the total number of sites along 
the $y$ axis
and $a_{0}$ the lattice constant.
Spinors are located at 
$X=\sqrt{3}a_{0}i$,
$i=1,\cdots, N_{X}$ within a unit cell
with $N_{X}$ being the total number of sites along the $X$ axis.
Hopping matrix elements for the zigzag case are given by
$
T_{+}^{}=(-t)\left[
\begin{array}{cc}
0  & 1 \\
0  & 0 
\end{array}\right],
$
$
T_{-}^{}=(-t)\left[
\begin{array}{cc}
0  & 0 \\
1  & 0 
\end{array}
\right],
$
and
$
T_{0}^{}=(-t)\left[
\begin{array}{cc}
0 & 1 \\
1 & 0
\end{array}
\right],
$
where $t$ is the hopping integral.
As specified by the chiral vector,
the periodic boundary condition is imposed along the edges,
which renders the wave numbers along the edges quantized,
$
k_{y}a_{0}= 2m \pi  /N_{y},m\in \mathbb{Z}
$.
The band structure is then composed of a set of 1D modes, 
each of which is characterized by the wave number.
Performing
a Fourier transformation along the $y$ axis as
$
\boldsymbol{c}(i,j)=1/\sqrt{N_{y}}\sum_{k_{y}}e^{\mathrm{i}k_{y}y}
\boldsymbol{c}_{k_{y}}(i)
$,
$
H_{\hbox{\tiny{kin}}}
$
is decomposed as
$
H_{\hbox{\tiny{kin}}}=\sum_{k_{y}}H(k_{y})
$,
\begin{eqnarray}
H(k_{y})
&=&\sum_{i=1}^{N_{X}}
\Big[
{\boldsymbol c}_{k_{y}}^{\dagger}(i)\,
V_{X}^{}\,
{\boldsymbol c}_{k_{y}}^{\vphantom{\dagger}}(i+1)+h.c.
\nonumber \\
&&
+{\boldsymbol c}^{\dagger}_{k_{y}}(i)\,
V_{0}^{}\,
{\bf c}_{k_{y}}^{\vphantom{\dagger}}(i)
\Big],
\end{eqnarray}
where
$
V_{X}^{}=
T_{-}^{}e^{-\mathrm{i}k_{y}a_{0}}
$
and
$
V_{0}^{}=T_{0}^{}+T_{-}^{}e^{+\mathrm{i}k_{y}a_{0}}
+T_{+}^{}e^{-\mathrm{i}k_{y}a_{0}}
$.
This 1D Hamiltonian can be seen as a 1D chain with 
alternating hoppings.\cite{lin98}
As for bulk electron states,
the energy spectrum is gapless for 
two values of $k_{y}$,
$k_{y}a_{0}=\pm k_{0}a_{0}=\pm 2\pi /3$,
whereas it is gapped for other $k_{y}$.
This is best seen by performing a gauge transformation
$
c_{\bullet}(i) \to
(-)^{i}e^{+\mathrm{i}k_{y}a_{0}/4}c_{\bullet}(i)
$,
$
c_{\circ}(i) \to 
(-)^{i}e^{-\mathrm{i}k_{y}a_{0}/4}c_{\circ}(i)
$,
which transforms the Hamiltonian as
$
V_{X}^{}\to
(-e^{-\mathrm{i} \frac{3}{2} k_{y}a_{0}})T_{-}
$
and 
$
V_{0}^{}\to
2\cos(k_{y}a_{0}/2)T_{0}
$.
Then,
the alternation can be completely removed for $k_{y}=\pm k_{0}$,
where the Hamiltonian is reduced to
$
H(\pm k_{0})
=
-t\sum_{l=1}^{N_{x}}
\big[
c^{\dagger}_{\pm,l}
c_{\pm,l+1}^{}+h.c.
\big]
$
by defining
$
c_{\pm k_{0},\bullet}(i)=:c_{\pm,l=2i-1}
$,
$
c_{\pm k_{0},\circ}(i)=:c_{\pm,l=2i}
$,
and 
$2N_{X}=:N_{x}$.

When truncated with edges,
the above one-parameter family of Hamiltonians 
supports zero-energy boundary states 
for some specific values of $k_{y}$,
\cite{fujita96,ryu02}
as a manifestation of a nontrivial bulk band structure.
For zigzag edges,
they appear for
$-\pi < k_{y}a_{0} < -2\pi/3,+2\pi/3 < k_{y}a_{0}< +\pi$,
within a bulk energy gap 
at the Fermi energy.
The localization length of the boundary states
continuously increases as we vary $k_{y}a_{0}$ from $+\pi(-\pi)$
to $+2\pi/3 (-2\pi/3)$. 
For $k_{y}a_{0}=\pm \pi$, the boundary states is 
completely localized at the boundaries $i=1$ and $N_{X}$,
while the localization length for $k_{y}a_{0}=\pm 2\pi/3$
is infinite,
at which the branch of the localized boundary states 
continuously merges with 
the extended bulk states.
If we explicitly 
construct wave functions for the boundary states,
they are given by
\begin{eqnarray}
f_{k_{y},\bullet}(i)\propto
\big(-2e^{\mathrm{i} 3 k_{y} a_{0}/2} \cos k_{y}a_{0}/2 
\big)^{i-1},\quad
f_{k_{y},\circ}(i)=0
\nonumber \\
\label{wfn_zigzag}
\end{eqnarray}
($-\pi < k_{y}a_{0} < -2\pi/3,
   +2\pi/3 < k_{y}a_{0} < +\pi$),
after the gauge transformation.

\subsubsection{Bearded edges}

We turn to a CNT with bearded edges.
Atomic configurations
realizing bearded edges for 
$\pi$ electrons
were recently proposed 
based on an \textit{ab initio} calculation.
\cite{kusakabe02}
We choose a different unit cell from that for the zigzag case
as indicated in Fig. \ref{cnt}(b).
($X$ is now equal to $2\sqrt{3}a_{0}i$,
$i=1,\cdots,N_{X}$.)
We obtain a set of Hamiltonian parametrized by $k_{y}$
as in the zigzag case
with the hopping matrices $V_{X}$ and $V_{0}$ being given by
$
V_{X}^{}=T_{-}^{}+T_{-}^{}e^{-\mathrm{i}k_{y}a_{0}}
$,
and
$
V_{0}=T_{0}
$.
Again,
the bulk energy spectrum is gapless for 
$k_{y}=\pm k_{0}$,
for which
the Hamiltonian can be transformed to a simple form
$
H(\pm k_{0})
=
-t\sum_{l=1}^{N_{x}}
\big[
c^{\dagger}_{\pm,l}
c_{\pm,l+1}^{}+h.c.
\big]
$
by a gauge transformation
$
c_{\bullet}(i) \to
e^{+\mathrm{i}k_{y}a_{0}i/2}c_{\bullet}(i)
$,
$
c_{\circ}(i) \to 
e^{+\mathrm{i}k_{y}a_{0}i/2}c_{\circ}(i)
$.
On the other hand,
boundary states appear for different values of $k_{y}$
from the zigzag case,
$-2\pi/3 < k_{y}a_{0} < +2\pi/3$.
The wave function of the boundary states can be 
explicitly constructed as
\begin{eqnarray}
f_{k_{y},\bullet}(i)\propto
\big(-2\cos k_{y}a_{0}/2 \big)^{-(i-1)},\quad
f_{k_{y},\circ}(i)=0
\label{wfn_bearded}
\end{eqnarray}
($-2\pi/3 < k_{y}a_{0} < +2\pi/3$),
after the gauge transformations.

\subsection{Interactions}
\label{Interactions}

As the boundary states with different $k_{y}$
are all degenerate at the Fermi energy,
electron interactions have pronounced effects.
In what follows,
we will construct low-energy effective theories describing 
correlation effects on the boundary states.
The interacting part of the Hamiltonian is written as
\begin{eqnarray}
H_{\hbox{\tiny{int}}}&=&
\frac{1}{2}
\sum_{RR^{\prime};pp^{\prime};ss^{\prime}}
V_{RR^{\prime};pp^{\prime};ss^{\prime}}
\nonumber \\
&&
\times
c_{ps}^{\dagger}(R)
c_{p^{\prime}s^{\prime}}^{\dagger}(R^{\prime})
c_{p^{\prime}s^{\prime}}^{}(R^{\prime})
c_{ps}^{}(R),
\end{eqnarray}
where 
$R=(i,j)$ runs over the lattice cites in the tight-binding model,
$p=\bullet/\circ$ represents a sublattice,
and $s=\uparrow/\downarrow$ is a spin index.
As for $V_{RR^{\prime};pp^{\prime};ss^{\prime}}$,
we take either the Hubbard interaction
\begin{eqnarray}
V_{RR^{\prime};pp^{\prime};ss^{\prime}}
=U\delta_{RR^{\prime}}\delta_{pp^{\prime}}\delta_{s,-s^{\prime}}
\end{eqnarray}
or the unscreened Coulomb interaction
\begin{eqnarray}
V_{RR^{\prime};pp^{\prime};ss^{\prime}}
=
\frac{e^{2}/\kappa}{
\sqrt{
(x_{p}-x^{\prime}_{p^{\prime}})^{2}
+4R^{2}\sin^{2} 
\left(\frac{y-y^{\prime}}{2R}\right)
+r_{z}^{2}
}
},
\nonumber \\
\end{eqnarray}
where $\kappa$ is an effective dielectric constant of the system,
$x_{\bullet}=X$, 
$x_{\circ}=X+\sqrt{3}a_{0}/2$ (zigzag),
$\sqrt{3}a_{0}$ (bearded),
$R$ is the tube radius,
and $r_{z}\sim a_{0}$ characterizes the radius of $p_{z}$ orbital,
serving as a short-distance cutoff.
In 
Ref. \onlinecite{egger97},
$\kappa$ is estimated to $\sim 1.4$,
while 
$r_{z}$ is determined to be $0.526\times a_{0}$
in Ref. \onlinecite{odintsov99},
from the requirement that the on-site interaction
in the original tight-binding model corresponds 
to the difference between the ionization potential 
and electron affinity of $sp^{2}$ hybridized carbon.
Since the Coulomb interaction is unscreened in CNT's,
the Hubbard interaction is less realistic.
However, as we will see, their differences 
are small for the correlation effects at boundaries,
due to the special properties of the boundary states,
although
the long-range nature of the Coulomb interaction
has fundamental effects for 
the bulk electron states.

Focusing on a low-energy sector,
we construct a effective theory
with boundary states at zero energy.
When a CNT is metallic,
two 1D gapless modes are also included,
while we drop all the operators belonging to gapped bands:
\begin{eqnarray}
\boldsymbol{c}(i,j)&=&
\frac{1}{\sqrt{N_{y}}}
\sum_{k_{y}}
e^{\mathrm{i}k_{y}y}\boldsymbol{c}_{k_{y}}(i)
\nonumber \\
&\longrightarrow&
\frac{1}{\sqrt{N_{y}}}
\sum_{\alpha=\pm}
e^{\mathrm{i}\alpha k_{0}y}
\boldsymbol{c}_{\alpha}(i)
+
\frac{1}{\sqrt{N_{y}}}
\boldsymbol{b}(i,j),
\nonumber \\
\label{dropping_operators}
\end{eqnarray}
where $\boldsymbol{c}_{\pm}(i):=\boldsymbol{c}_{\pm k_{0}}(i)$,
$\boldsymbol{b}$
is a linear combination of boundary states,
$
\boldsymbol{b}(i,j)=
\sum_{k_{y}}^{\prime}e^{\mathrm{i}k_{y}y}\boldsymbol{f}_{k_{y}}(i)\,e_{k_{y}}
$,
with $\boldsymbol{f}_{k_{y}}(i)$ being an eigen wave function localized at boundaries,
$e_{k_{y}}^{\dagger}$
and $e_{k_{y}}^{\vphantom{\dagger}}$
are creation and annihilation operators for a boundary state,
and $\sum^{\prime}$ means the summation is restricted to the wave numbers 
for which a boundary state appears.
We project the interaction $H_{\hbox{\tiny{int}}}$ 
into the reduced Hilbert space spanned by 
the two gapless modes 
$\boldsymbol{c}_{\pm}^{\dagger}$,
$\boldsymbol{c}^{ }_{\pm}$ 
and the boundary states 
$e_{k_{y}}^{\dagger}$,
$e_{k_{y}}^{}$.
Substituting Eq. (\ref{dropping_operators}),
$H_{\hbox{\tiny{int}}}$ is decomposed into a number of terms,
each of which is a four-fermion interaction made of either
$\boldsymbol{c}_{\pm}^{\dagger}$,
$\boldsymbol{c}_{\pm}^{}$,
$e_{k_{y}}^{\dagger}$, or 
$e_{k_{y}}^{}$.
However, not all of them give rise to a contribution
due to the momentum conservation along the edges.
For example, a four-fermion interaction 
composed of three bulk electrons and one edge electron
(e.g., 
$c_{+s}^{\dagger}c_{-s^{\prime}}^{\dagger}
c_{-s^{\prime}}^{\vphantom{\prime}}b_{s^{\prime}}^{\vphantom{\prime}}
$)
does not appear for the zigzag cases
since the momentum carried by the bulk electron
is either $+2\pi/3$ or $-2\pi /3$, 
while that carried by the edge electrons is never equal to 
$\pm 2\pi/3$,
and hence a momentum mismatch occurs.
It should be also noted that
boundary states are nonvanishing only on one of sublattices
$p=\bullet$
as seen from Eqs. (\ref{wfn_zigzag}) and (\ref{wfn_bearded}).
Due to this special property of boundary states
and also since the boundary states are localized at boundaries,
the Coulomb interaction and the Hubbard interaction
are not quite different.
So we restricted ourselves to the Hubbard interaction
for a while
as it is simpler.
We write the projected interaction as
$H_{\hbox{\tiny{int}}}\rightarrow
H_{\hbox{\tiny{int}}}^{\hbox{\tiny{bulk}}}
+H_{\hbox{\tiny{int}}}^{\hbox{\tiny{edge}}}
$,
where $H_{\hbox{\tiny{int}}}^{\hbox{\tiny{bulk}}}$
is solely composed of bulk electron operators,
while $H_{\hbox{\tiny{int}}}^{\hbox{\tiny{edge}}}$ includes
boundary states.
We further decompose $H_{\hbox{\tiny{int}}}^{\hbox{\tiny{edge}}}$ 
as
$H_{\hbox{\tiny{int}}}^{\hbox{\tiny{edge}}}=
H_{\hbox{\tiny{int}}}^{\hbox{\tiny{edge4}}}+
H_{\hbox{\tiny{int}}}^{\hbox{\tiny{edge3}}}+
H_{\hbox{\tiny{int}}}^{\hbox{\tiny{edge2}}}+
H_{\hbox{\tiny{int}}}^{\hbox{\tiny{edge1}}}
$.
Firs,
$H_{\hbox{\tiny{int}}}^{\hbox{\tiny{edge4}}}$
consists solely of boundary states,
\begin{eqnarray}
H_{\hbox{\tiny{int}}}^{\hbox{\tiny{edge4}}}
&=&
\frac{U}{N_{y}}
\sum_{k_{1},k_{2},k_{3},k_{4}}^{\prime}
\delta(-k_{1}+k_{2}-k_{3}+k_{4})
\nonumber \\
&&
\times
e_{k_{1}\uparrow}^{\dagger}e_{k_{2}\uparrow}^{\vphantom{\dagger}}
e_{k_{3}\downarrow}^{\dagger}e_{k_{4}\downarrow}^{\vphantom{\dagger}}
\sum_{i}
f_{k_{1}}^{*}(i)
f_{k_{2}}^{\vphantom{*}}(i)
f^{*}_{k_{3}}(i)
f_{k_{4}}^{\vphantom{*}}(i),
\nonumber \\
\label{H_edge4}
\end{eqnarray}
where we simply wrote 
$f_{k,p=\bullet}$ as $f_{k}$
and $\delta(k)$ is a lattice delta function;
i.e., $\delta(k)$ is equal to unity only when $k$ 
is integral multiple of $2\pi$,
while it is vanishing otherwise.
When a CNT is insulating,
it is enough to consider only $H_{\hbox{\tiny{int}}}^{\hbox{\tiny{edge4}}}$ 
for boundary physics.
Then, the problem is reduced to 
how the degeneracy between the boundary states
with quenched kinetic energy
is lifted by $H_{\hbox{\tiny{int}}}^{\hbox{\tiny{edge4}}}$,
as in the flatband magnetism
or the fractional quantum Hall effect.
Note also that
the strength of the Hubbard interaction 
is reduced by the factor $1/N_{y}$,
as the wave functions of the boundary states are extended 
over the circumference of the tube.

The part composed of two edge operators
$H_{\hbox{\tiny{int}}}^{\hbox{\tiny{edge2}}}$
is given by
\begin{eqnarray}
H_{\hbox{\tiny{int}}}^{\hbox{\tiny{edge2}}}
&=&
\frac{U}{N_{y}}
\sum_{i}
\Big[
-2
\tilde{J}^{z}
\tilde{S}^{z}
-
\tilde{J}^{+}
\tilde{S}^{-}
-
\tilde{S}^{+}
\tilde{J}^{-}
\nonumber \\
&&
+\frac{1}{2}
\tilde{\rho}^{\vphantom{\dagger}}
\tilde{\rho}_{e}^{\vphantom{\dagger}}
+
\tilde{\Delta}^{\dagger}
\tilde{\Delta}_{e}^{\vphantom{\dagger}}
+
\tilde{\Delta}_{e}^{\dagger}
\tilde{\Delta}^{\vphantom{\dagger}}
\Big],
\label{H_edge2}
\end{eqnarray}
where
\begin{eqnarray}
&&
%\tilde{J}^{a}=\sum_{\alpha,ss^{\prime}}
\tilde{\boldsymbol{J}}=\sum_{\alpha,ss^{\prime}}
c_{\alpha s}^{\dagger}
\frac{\boldsymbol{\sigma}_{ss^{\prime}}}{2}
c_{\alpha s^{\prime}}^{},
\quad
%\tilde{S}^{a}=
\tilde{\boldsymbol{S}}=
\sum_{k_{y},ss^{\prime}}^{\prime}
|f_{k_{y}}|^{2}
e^{\dagger}_{k_{y}s}
\frac{\boldsymbol{\sigma}_{ss^{\prime}}}{2}
e^{}_{k_{y}s^{\prime}},
\nonumber \\
&&
\tilde{\rho}=
\sum_{\alpha,s}c_{\alpha s}^{\dagger}c_{\alpha s}^{},
\qquad
\tilde{\rho}_{e}
=
\sum_{k_{y},s}^{\prime}
|f_{k_{y}}|^{2}
e^{\dagger}_{k_{y}s}e^{}_{k_{y}s},
\nonumber \\
&&
\tilde{\Delta}=
\Big[
c_{+\downarrow}c_{-\uparrow}
+
c_{-\downarrow}c_{+\uparrow}
\Big],\quad
\tilde{\Delta}_{e}
=
\sum_{k_{y}}^{\prime}
f_{k_{y}}^{2}
e^{}_{k_{y}\uparrow}e^{}_{-k_{y}\downarrow}.
\nonumber \\
\label{def_currents}
\end{eqnarray}
We omit sublattice indices and simply write
$c_{\pm}=c_{\pm \bullet}$,
$b=b_{\bullet}$ henceforth,
since $p=\circ$ does not appear.
Finally,
$H_{\hbox{\tiny{int}}}^{\hbox{\tiny{edge1}}}$
and 
$H_{\hbox{\tiny{int}}}^{\hbox{\tiny{edge3}}}$
are given by
\begin{eqnarray}
H_{\hbox{\tiny{int}}}^{\hbox{\tiny{edge1}}}
&=&
\frac{U}{N_{y}^{2}}\sum_{i,j}
\Big[
c^{\dagger}_{+\uparrow}c^{\vphantom{\dagger}}_{-\uparrow}
\left(
c^{\dagger}_{+\downarrow}b^{\vphantom{\dagger}}_{-\downarrow}
+
b^{\dagger}_{\downarrow}c^{\vphantom{\dagger}}_{+\downarrow}
\right)
+h.c.
\nonumber \\
&&
+
\left(
c^{\dagger}_{+\uparrow}b^{\vphantom{\dagger}}_{-\uparrow}
+
b^{\dagger}_{\uparrow}c^{\vphantom{\dagger}}_{+\uparrow}
\right)
c^{\dagger}_{+\downarrow}c^{\vphantom{\dagger}}_{-\downarrow}
+
h.c.
\Big],
\nonumber \\
H_{\hbox{\tiny{int}}}^{\hbox{\tiny{edge3}}}
&=&
\frac{U}{N_{y}^{2}}\sum_{i,j}
\Big[
b^{\dagger}_{\uparrow}b^{\vphantom{\dagger}}_{\uparrow}
\big(
c_{-\downarrow}^{\dagger}b_{\downarrow}^{\vphantom{\dagger}}
+b^{\dagger}_{\downarrow}c_{+\downarrow}^{\vphantom{\dagger}}
\big)
+
h.c.
\nonumber \\
&&
+
\big(
c_{-\uparrow}^{\dagger}b_{\uparrow}^{\vphantom{\dagger}}
+b^{\dagger}_{\uparrow}c_{+\uparrow}^{\vphantom{\dagger}}
\big)
b^{\dagger}_{\downarrow}b_{\downarrow}^{\vphantom{\dagger}}
+
h.c.
\Big].
\end{eqnarray}
As commented above, $H_{\hbox{\tiny{int}}}^{\hbox{\tiny{edge1}}}$
does not appear for the zigzag cases
due to a momentum mis-match.
Furthermore,
for a thin CNT with zigzag edges,
$H_{\hbox{\tiny{int}}}^{\hbox{\tiny{edge3}}}$
is also vanishing 
and hence 
the effective Hamiltonian is simplified.

The bulk part of a metallic CNT is described by a
simple tight-binding Hamiltonian
$
H(\pm k_{0})
=
-t\sum_{l=1}^{N_{x}}
\big[
c_{\pm,l}^{\dagger}c_{\pm,l+1}^{}+h.c.
\big]
$
after the gauge transformation.
In describing the gapless excitations,
we replace lattice fermion operators $c_{\pm,l}$by
slowly varying continuum operators $\psi_{L/R,\pm}(x)$ as
\begin{eqnarray}
c_{\pm,l}\sim
\sqrt{a_{x}}
\Big[
e^{+\mathrm{i}k_{F}x}\psi_{L,\pm}(x)
+
e^{-\mathrm{i}k_{F}x}\psi_{R,\pm}(x)
\Big],
\end{eqnarray}
where $a_{x}:=\sqrt{3}a_{0}/2$,
and $x:=2X$.
Due to the open boundary conditions, however,
the left and right movers are not independent.
They satisfy a constraint
\begin{eqnarray}
\psi_{R,\pm}(x)=-\psi_{L,\pm}(-x),
\label{constraint}
\end{eqnarray}
which allows us
to concentrate on 
only the left-moving sector, say.
Focusing on the two gapless modes,
the kinetic part of the system is written by the slowly varying 
variables as
\begin{eqnarray}
H_{\hbox{\tiny{kin}}}= v_{F}\sum_{\alpha,s}\int_{-L}^{+L} dx\,
\psi^{\dagger}_{L\alpha s} \mathrm{i}\partial_{x}\psi^{}_{L\alpha s},
\label{H_kin_psi}
\end{eqnarray}
where 
$L= N_{x} a_{x}$
and
$v_{F}=\sqrt{3}ta_{0}/2$ is the Fermi velocity.
Here, the original system defined for $x\in [0,+L]$
is extended to $x\in [-L,+L]$ by the constraint,
Eq. (\ref{constraint}).

Upon taking continuum limit, 
we are also allowed to set $X=x=0$
for the interactions that include boundary states
with some renormalizations of couplings,
since edge modes are exponentially localized at the boundary.
The lattice fermion operators at the 
boundary are replaced with the left moving continuum field as
\begin{eqnarray}
c_{\pm,l=1}
&\rightarrow&
\sqrt{a_{x}}
\Big[
e^{+\mathrm{i}k_{F}a_{x}}\psi_{L,\pm}(a_{x})
+
e^{-\mathrm{i}k_{F}a_{x}}\psi_{R,\pm}(a_{x})
\Big]
\nonumber \\
&\sim&
\sqrt{a_{x}}\,
2\mathrm{i}\sin (k_{F}a_{x}) \,
\psi_{L,\pm}(0).
\end{eqnarray}
Correspondingly,
operators made of lattice fermions
$\tilde{\boldsymbol{J}},\tilde{\rho},\tilde{\Delta}^{\dagger},
\tilde{\Delta}$
appearing in $H^{\hbox{\tiny{edge2}}}_{\hbox{\tiny{int}}}$ 
are replaced with
continuum counterparts as
\begin{eqnarray}
&&
\boldsymbol{J}=\sum_{\alpha,ss^{\prime}}
\psi_{L\alpha s}^{\dagger}
\frac{\boldsymbol{\sigma}_{ss^{\prime}}}{2}
\psi_{L\alpha s^{\prime}}^{},\quad
\rho=
\sum_{\alpha,s}\psi_{L\alpha s}^{\dagger}\psi_{L\alpha s}^{},
\nonumber \\
&&
\Delta=
\Big[
\psi_{L+\downarrow}\psi_{L-\uparrow}
+
\psi_{L-\downarrow}\psi_{L+\uparrow}
\Big],
\label{def_currents_cont}
\end{eqnarray}
with suitable renormalizations for couplings.
At this level, also, 
differences between the Hubbard and the Coulomb interaction
are irrelevant,
since we keep the same terms for the both types of interactions.

\subsection{Effects of the bulk interactions:
Open boundary bosonization}
\label{Effects of the bulk interactions:
Open boundary bosonization}

Although we are interested in physics at boundaries,
effects of electron correlation in the bulk regime
$H_{\hbox{\tiny{int}}}^{\hbox{\tiny{bulk}}}$
should also be taken into account,
since the bulk interaction is known to affect 
the scaling dimension of operators inserted at a boundary
in 1D correlated systems.
The bulk interaction is well incorporated by
the bosonization technique
for 1D systems.
Following Ref. \onlinecite{fabrizio95},
we bosonize the theory with open boundary condition.
Open boundary bosonization was previously used 
to discuss the correlation effects of 
armchair CNTs with boundaries,
where there are no boundary states though.
\cite{egger98,yoshioka02,yoshioka02-2}

Upon bosonization,
electron operators are expressed in terms of 
scalar bosonic operators as
\begin{eqnarray}
\psi_{\alpha s}(x)\equiv\frac{1}{\sqrt{2\pi a_{x}}}\eta_{\alpha s}
:e^{+\mathrm{i}\sqrt{4\pi}\varphi_{\alpha s}}:(x),
\label{mandelstam}
\end{eqnarray}
where $\eta_{\alpha s}$ is a Klein factor,
and $:\cdots:$ denotes normal ordering.
[Since it is enough to focus on the holomorphic sector,
we simply write $\psi_{L\alpha s}(x)=\psi_{\alpha s}(x)$,
$\varphi_{L\alpha s}(x)=\varphi_{\alpha s}(x)$, etc.,
henceforth.]
For convenience, it is better to introduce 
a new basis 
$\left\{\varphi_{\rho,\pm},\varphi_{\sigma,\pm}\right\}$
rather than $\varphi_{\pm,\uparrow/\downarrow}$,
\begin{eqnarray}
\varphi_{\alpha s}(x)
&:=&\frac{1}{2}
\Big\{
\varphi_{\rho +}(x)
+s\varphi_{\sigma +}(x)
\nonumber \\
&&
+\alpha \varphi_{\rho -}(x)
+\alpha s\varphi_{\sigma -}(x)
\Big\}.
\end{eqnarray}

When the effects of forward scatterings are taken into account,
left and right movers are mixed through the Bogoliubov transformation.
Since there exists the constraint between the left and right movers,
bosonization rules take non-local form,
where $\varphi_{\alpha s}$ in Eq. (\ref{mandelstam})
is,
after the Bogoliubov transformation,
given by
\begin{eqnarray}
\varphi_{j\delta}(x)
&\rightarrow &
\frac{1}{2}
\Big\{
\sqrt{K_{j\delta}}
\left[
\varphi_{j\delta}(+x)-
\varphi_{j\delta}(-x)
\right]
\nonumber \\
&&
+
\frac{1}{\sqrt{K_{j\delta}}}
\left[
\varphi_{j\delta}(+x)+
\varphi_{j\delta}(-x)
\right]
\Big\},
\end{eqnarray}
where $j=\rho/\sigma$ and $\delta=+/-$.
Especially, at the boundary $x=0$,
\begin{eqnarray}
\varphi_{ \alpha s}(0)
&=&
\frac{1}{2}
\Big\{
\varphi_{\rho +}(0)/\sqrt{K_{\rho+}}
+s
\varphi_{\sigma +}(0)/\sqrt{K_{\sigma +}}
\nonumber \\
&&
+\alpha
\varphi_{\rho -}(0)/\sqrt{K_{\rho -}}
+\alpha s
\varphi_{\sigma -}(0)/\sqrt{K_{\sigma -}}
\Big\}.
\nonumber \\
\end{eqnarray}
Parameters
$K_{j\delta}$ are the Luttinger parameters for each mode
--
that is, a coefficient of the Bogoliubov transformation.
The long-range bulk forward scattering 
strongly renormalizes the charge symmetric mode $(\rho,+)$,
and its Luttinger parameter
is estimated to be $K_{\rho+}\sim 0.2$
for a CNT with the Coulomb interaction.
\cite{kane97}
On the other hand, for the other  modes,
$K_{j\delta}$ is almost equal to unity
if we neglect the back scatterings and the umklapp scattering.

The bulk interactions affect physics at boundaries
through the modifications of the Luttinger parameters.
More precisely, they alter the scaling dimensions of 
the operators inserted at boundaries.
The scaling dimensions of 
$J^{\pm}$ and $\Delta^{\dagger},\Delta$
in Eq. (\ref{def_currents_cont}),
which we call
$x_{\perp}$ and $x_{\Delta}$,
respectively,
are equal to unity in the absence of the interactions.
However,
they are now given by
\begin{eqnarray}
x_{\perp}=
\frac{
K_{\sigma+}^{-1}
+
K_{\sigma-}^{-1}
}{2},\quad
x_{\Delta}=
\frac{
K_{\rho+}^{-1}
+
K_{\sigma-}^{-1}
}{2},
\end{eqnarray}
which are dependent on the Luttinger parameters.
We see that
superconducting pairing operators $\Delta^{\dagger},\Delta$ 
are made to be strongly irrelevant
by the strong bulk repulsive interactions.

The bosonized expression for
the bulk part of the Hamiltonian is
\begin{eqnarray}
H_{\hbox{\tiny{kin}}}+
H_{\hbox{\tiny{int}}}^{\hbox{\tiny{}bulk}}&=&
H_{0}^{\hbox{\tiny{}bulk}}+
H_{\hbox{\tiny{int}}}^{\hbox{\tiny{bulk}}\prime},
\nonumber \\
H_{0}^{\hbox{\tiny{}bulk}}
&=&
\sum_{j\delta} \pi v_{j\delta}\int_{-L}^{+L} dx:J_{j\delta}(x)^{2}:,
\end{eqnarray}
where
$J_{j\delta}=-\partial_{x}\varphi_{j\delta}/\sqrt{\pi}$
represents a $U(1)$-current
and $v_{j\delta}$ is a velocity for each collective mode.
The velocity for the charge symmetric mode is
strongly enhanced as $v_{\rho+}\sim v_{F}/K_{\rho+}$,
while the velocity is almost equal to the Fermi velocity 
for other modes,
$v_{j\delta}\sim v_{F}$,
$(j,\delta)\neq (\rho,+)$.
$H_{\hbox{\tiny{int}}}^{\hbox{\tiny{bulk}}\prime}$
represents the part that cannot 
be written as a current-current interaction,
in the presence of which
the modes $\sigma\pm$ and $\rho -$ are made to be gapped away from half filling,
whereas 
all kinds of excitations are gapped
at half filling 
due to the umklapp scatterings.
\cite{egger97,egger98,yoshioka99}

\section{Case of one boundary state}
\label{Case of one boundary state}

Having established low-energy effective theories
that describe correlation effects at boundaries,
we now turn to specific examples.
We first consider a thin metallic CNT with
only one boundary state,
such as
$(6,-6)$ CNT with zigzag edges or
$(3,-3)$ CNT with bearded edges.
Especially, we focus on the former example
near half filling
as it is the simplest case in that
its boundary state $\boldsymbol{f}_{\pi}(i)$ 
appearing for $k_{y}a_{0}=\pi$
is completely localized at boundaries,
$f_{\pi,p}(i)=\delta_{i,1}\delta_{p,\bullet}$
[Eq. (\ref{wfn_zigzag})]
.

The most general expression 
for low-energy effective theories 
is now reduced to
the following Hamiltonian:
\begin{eqnarray}
H&=&H_{0}^{\hbox{\tiny{bulk}}}
+H_{\hbox{\tiny{int}}}^{\hbox{\tiny{bulk}}\prime}
+H_{\hbox{\tiny{int}}}^{\hbox{\tiny{edge}}}
=:H_{0}+H_{I}+H_{\hbox{\tiny{int}}}^{\hbox{\tiny{bulk}}\prime},
\nonumber \\
H_{0}&=&
\sum_{j\delta} \pi v_{j\delta}\int_{-L}^{+L} dx:J_{j\delta}(x)^{2}:
+\epsilon_{e}\rho_{e}
+U_{e}n_{\uparrow}n_{\downarrow},
\nonumber \\
H_{I}&=&
\frac{v_{F}\lambda_{\rho}}{4}\,\rho_{e} \, \rho 
+v_{F}\lambda_{z}\, S^{z}\, J^{z}
\nonumber \\
&&
+\frac{v_{F}\lambda_{\perp}}{2}
(v_{F}\tau_{c})^{x_{\perp}-1}
\Big[
S^{+}J^{-}+J^{+}S^{-}
\Big],
\end{eqnarray}
where 
$\boldsymbol{J},\rho$
are defined in Eq. (\ref{def_currents_cont}).
Superconducting pairing operators
$\Delta^{\dagger},\Delta$
are dropped as they are strongly irrelevant.
Also,
$\tilde{\boldsymbol{S}}$ and $\tilde{\rho}_{e}$
are reduced to 
$
\boldsymbol{S}
=\sum_{ss^{\prime}}
e_{s}^{\dagger}\,
\boldsymbol{\sigma}_{ss^{\prime}}/2\,
e_{s^{\prime}}^{}
$
and
$
\rho_{e}=\sum_{s}e_{s}^{\dagger}e_{s}^{}
$,
respectively.
An ultraviolet cutoff $\tau_{c}$
is introduced,
which is estimated to be the inverse of the bandwidth,
$\tau_{c}\sim 1/t$
($v_{F}\tau_{c}\sim a_{x}$).
$\lambda_{\rho,z,\perp}$ represents a dimensionless coupling
between the boundary states and conduction electrons.
For a $(6,-6)$ CNT with zigzag edges,
initial conditions for RG analysis
(bare values of the couplings)
are given by
$U_{e}=U/N$,
$\epsilon_{e}\sim 0$,
and
$-v_{F}\lambda_{z}/2
=-v_{F}\lambda_{\perp}/2
=v_{F}\lambda_{\rho}/2
%=v_{F}\lambda_{\Delta}
=4a_{x}\sin^{2}(k_{F}a_{x})U/N$
%$\lambda_{U}=\lambda_{\epsilon}=0$,
for the Hubbard interaction
near half filling,
whereas they are given by
$U_{e}=\tilde{V}(0)$,
$\epsilon_{e}\sim 0$,
$
-v_{F}\lambda_{z}/2
=-v_{F}\lambda_{\perp}/2
%=v_{F}\lambda_{\Delta}
=4a_{x}\sin^{2}(k_{F}a_{x}) \tilde{V}(k_{0}+\pi)$,
%$\lambda_{U}=\lambda_{\epsilon}=0$,
and
$v_{F}\lambda_{\rho}/2=4a_{x}\sin^{2}(k_{F}a_{x})
\big[2\tilde{V}(0)-\tilde{V}(k_{0}+\pi)\big]$
for the Coulomb interaction,
where
$
\tilde{V}(q):=
N_{y}^{-1}\sum_{y_{j}}
e^{-\mathrm{i}qy_{j}}V(y_{j})
$
and
$
V(y_{j}-y_{j^{\prime}})=
V_{p=p^{\prime}=\bullet;i=i^{\prime}=1;j,j^{\prime}}
$.
Thus, if we switch off the interaction
between boundary states and conduction electrons ($H_{I}$),
the ground state at the boundary 
at (or slightly above) half filling
is the state where one of the two boundary states is occupied.

%\subsection{Tomonaga-Luttinger liquid regime}
%\label{Tomonaga-Luttinger liquid regime}

To see what happens when we switch on couplings between 
boundary and conduction electrons,
we perform a perturbative RG analysis up to one-loop order
\cite{si93}
via a Coulomb gas representation of the partition function,
with $H_{0}$ being an unperturbed part of the Hamiltonian.
We neglect $H_{\hbox{\tiny{int}}}^{\hbox{\tiny{bulk}}\prime}$
for the time being,
and consider the case where the system is described by TL liquid,
i.e., temperature above the gaps
induced by $H_{\hbox{\tiny{int}}}^{\hbox{\tiny{bulk}}\prime}$.
This is a good approximation for 
the systems off the half filled condition.
Infinitesimally rescaling the ultraviolet cutoff,
$\tau_{c}\rightarrow \tau_{c}e^{-dl}$,
we obtain a set of RG equations
\begin{eqnarray}
&&
\frac{d\lambda_{\rho}}{dl}
=0,
\qquad
\frac{d\lambda_{z}}{dl}
=
\frac{\lambda_{\perp}^{2}}{ \sqrt{K_{\sigma +}}},
\nonumber \\
&&
\frac{d\lambda_{\perp}}{dl}
=
(1-x_{\perp})\lambda_{\perp}
+
\frac{\lambda_{z}\lambda_{\perp}}{\sqrt{K_{\sigma+}}},
\nonumber \\
&&
\frac{dh_{\epsilon}}{dl}
=
h_{\epsilon}
-\left(
\frac{\lambda_{\rho}^{2}}{4}
+
\frac{\lambda_{z}^{2}}{4}
+
\frac{\lambda_{\perp}^{2}}{2}
\right),
\nonumber \\
&&
\frac{dh_{U}}{dl}
=
h_{U}
-
2
\left(
\frac{\lambda_{\rho}^{2}}{4}
-
\frac{\lambda_{z}^{2}}{4}
-
\frac{\lambda_{\perp}^{2}}{2}
\right),
\label{rg_one_edge}
\end{eqnarray}
where dimensionless couplings are defined as
$h_{\epsilon}:=\tau_{c}\epsilon_{e}$,
$h_{U}:=\tau_{c}U_{e}$,
%$h_{U+2\epsilon}:=\tau_{c}(U_{e}+2\epsilon_{e})$,
and we performed the rescaling
\begin{eqnarray}
&&
\lambda_{\rho}\rightarrow
\frac{\sqrt{K_{\rho+}}}{2\pi}
\frac{v_{F}}{v_{\rho+}}\lambda_{\rho},\quad
\lambda_{z}\rightarrow
\frac{\sqrt{K_{\sigma+}}}{2\pi}
\frac{v_{F}}{v_{\sigma+}}\lambda_{z},
\nonumber \\
&&
\lambda_{\perp}\rightarrow
\frac{1}{2\pi} 
\left[\frac{v_{F}}{v_{\sigma+}}\right]
^{\frac{1}{2K_{\sigma+}}}
\left[\frac{v_{F}}{v_{\sigma-}}\right]
^{\frac{1}{2K_{\sigma-}}} \lambda_{\perp},
\label{rescale}
\end{eqnarray}
to simplify the RG equations.
We can treat $\lambda_{z}$ and $\lambda_{\rho}$
non-perturbatively
in the manner of Schotte and Schotte
\cite{schotte69},
which actually gives 
an identical result 
to the above one-loop calculation
for $dh_{\epsilon}/dl$ and $dh_{U}/dl$.
However,
we prefer to treat
$\lambda_{z},\lambda_{\rho}$
and 
$\lambda_{\perp}$
on an equal footing here.

From the RG equations,
we see that interactions in the spin sector 
$\lambda_{z,\perp}$ are renormalized to zero
as it is well known that the ferromagnetic, isotropic Kondo interactions
are vanishing in the infrared.
The bulk electron correlations are almost irrelevant 
for the RG equations of the spin sector.
On the other hand,
the repulsive bulk interactions profoundly
affect the charge sector;
they suppress the charge fluctuations at boundaries.
First, as already commented,
the scaling dimensions of $\Delta^{\dagger}$
and $\Delta$,
which allow a pairwise hopping from the bulk part to 
boundary states and vice versa,
are made strongly irrelevant.
Second,
they modify the values of couplings through the Bogoliubov transformation
as seen from Eq. (\ref{rescale}).
When the bulk repulsive interaction is very strong $K_{\rho+}<< 1$,
the bare value of $\lambda_{\rho}$ is drastically reduced 
after the rescaling (\ref{rescale}),
which amounts to 
$U_{e} \to +\infty$
and 
$U_{e}+2\epsilon_{e} \to +\infty$
as $l\to +\infty$.
Then,
doubly occupying a boundary state
is prohibited in the infrared limit.
Note also that even though we start from 
slightly below half filling $h_{\epsilon} \gtrsim 0$,
for which no boundary states are occupied
if we switch off the coupling between bulk and boundary electrons,
couplings $\lambda_{\rho,z,\perp}$ renormalize $h_{\epsilon}$
to $-\infty$
and hence one of two 
boundary states is occupied in the ground state,
with the total spin carried by the boundary states being equal to $1/2$.
We numerically solve the RG equations (\ref{rg_one_edge})
for the Coulomb interaction
and confirmed that this is the case for the Coulomb interaction
\cite{rem1}.
%On the other hand,
%the effects of the Bogoliubov transformation
%are small for the spin sector
%as $K_{\sigma \pm}$ is nearly equal to unity.

At the fixed point
$\lambda_{z,\perp}\to 0$,
$\epsilon_{e}\to -\infty$,
and 
$U_{e},U_{e}+2\epsilon_{e} \to +\infty$,
the conduction electron 
is described as isolated TL liquid  
with the open boundary condition,
where tunneling density of states or 
Freidel oscillation 
can exhibit characteristic behavior
\cite{kane97,egger98,yoshioka02,yoshioka02-2}.
Correlation effects between bulk and boundary states
around this fixed point
can be taken into account in a perturbative way.

%\subsection{Infra-red regime}
%\label{Infra-red regime}

At very low temperature,
$H_{\hbox{\tiny{int}}}^{\hbox{\tiny{bulk}}\prime}$
causes 
energy gaps 
in the modes $\sigma\pm$ and $\rho-$ away from half filling,
whereas all kinds of excitations are gapped for half filling.
Effects of $H_{\hbox{\tiny{int}}}^{\hbox{\tiny{bulk}}\prime}$
on boundary physics are twofold.
First,
although we have treated the Luttinger parameters
as a fixed constant,
they undergo renormalizations
when the effects of $H_{\hbox{\tiny{int}}}^{\hbox{\tiny{bulk}}\prime}$
are taken into account.
The Luttinger parameters
$K_{\sigma+}$, $K_{\rho-}$, and $K_{\sigma-}^{-1}$ 
always renormalize to zero,
while $K_{\rho+}$ renormalizes to zero
only for half filling.\cite{egger97,egger98,yoshioka99}
This effect is easily accounted for
in the RG equations (\ref{rg_one_edge}),
which does not affect the conclusions
as $\Delta^{\dagger}$ and $\Delta$
stay irrelevant.
Second,
$H_{\hbox{\tiny{int}}}^{\hbox{\tiny{bulk}}\prime}$
might generate extra operators at the boundary
upon RG transformations.
Then, RG equations should be traced 
with these extra operators.
This is, however,
beyond the present discussion.

To get further insights,
it might be better to adopt
a complementary starting point 
rather than the above TL liquid model for higher temperature.
In the lower-temperature limit, slightly away from half filling,
a superconducting ground state
on the honeycomb lattice can be formed,
which originates from 
a rung-singlet state in the effective two-leg ladder model.
Emergence of isolated states at boundaries
can then be determined based on this
specific pattern of singlet pairs,
as in the valence-bond solid states in
spin systems.

\section{Case of two boundary states}
\label{Case of two boundary state}

As a next step,
we consider a thicker metallic CNT with 
two boundary states:
$(9,-9)$ CNT with zigzag edges,
for which 
boundary states appear 
for $k_{y}a_{0}=-8\pi/9$ and $+8\pi/9$.
(Fig. \ref{cnt})
Our main interest here is on
whether or not the total spin carried by the two boundary states
is non-zero.
The effective Hamiltonian 
for this case
is given by
\begin{eqnarray}
H_{0}&=&
\sum_{j\delta} \pi v_{j\delta}\int_{-L}^{+L} dx:J_{j\delta}(x)^{2}:
\nonumber \\
&&
+\frac{I}{4}\rho_{1}\rho_{2}
+K_{z}S_{1}^{z}S_{2}^{z}
+\frac{K_{\perp}}{2}
\Big[
S_{1}^{+}S_{2}^{-}+S_{2}^{+}S_{1}^{-}
\Big]
\nonumber \\
&&
+
U_{e}\Big[
n_{1\uparrow}n_{1\downarrow}
+n_{2\uparrow}n_{2\downarrow}
\Big]
+\epsilon_{e}\rho_{e},
\nonumber \\
H_{I}
&=&
\frac{v_{F}\lambda_{\rho}}{4}\rho_{e}\, \rho
+
v_{F}\lambda_{z}S^{z}J^{z}
\nonumber \\
&&
+
\frac{v_{F}\lambda_{\perp}}{2}
(v_{F}\tau_{c})^{x_{\perp}-1}
\Big[S^{+}J^{-}+J^{+}S^{-}\Big],
\end{eqnarray}
where we simply write 
$e_{1}:=e_{+8\pi/9}$,
$e_{2}:=e_{-8\pi/9}$.
Again,
superconducting pairing operators
$\Delta^{\dagger}$,$\Delta$ 
are dropped as they are irrelevant.
Initial conditions are given by
$K_{z}=K_{\perp}=-I=-2U_{e}<0$,
$\epsilon_{e}\sim 0$,
and
$\lambda_{z}=\lambda_{\perp}\sim -\lambda_{\rho}<0$.
Then, if we neglect the couplings between
conduction electrons and boundary states,
%($H_{I}$),
the ground state for the boundary 
near half filling
is found to be the state with 
the total spin equal to unity.

To see the effects of $H_{I}$,
RG equations for the couplings 
are obtained in the same way as the case
of one boundary state.
Again, we focus on TL liquid regime
and 
neglect $H_{\hbox{\tiny{int}}}^{\hbox{\tiny{bulk}}\prime}$.
RG equations for $\lambda_{z}$,
$\lambda_{\perp}$,
$h_{\epsilon}$, and 
$h_{U}$
are identical to those in the case of one boundary 
state.
Then, $\lambda_{z}$ and $\lambda_{\perp}$
become vanishing in the infrared limit,
since they are initially ferromagnetic and isotropic.
In addition,
charge fluctuations are suppressed when
there are the strong repulsive interactions in the bulk,
since $\epsilon_{e}\rightarrow-\infty$ and 
$U_{e},U_{e}+2\epsilon_{e}\rightarrow +\infty$,
and hence
doubly occupying a boundary is unfavorable.
RG equations for $K_{z,\perp}$ and $I$,
which determine the total spin 
carried by the ground state of $e$ fermions,
are given by
\begin{eqnarray}
&&
\frac{d h_{I}}{dl}
=
h_{I}
-2\lambda_{\rho}^{2},
\nonumber \\
&&
\frac{d h_{K_{z}}}{dl}
=
h_{K_{z}}
-
\lambda_{z}^{2},\quad
\frac{d h_{K_{\perp}}}{dl}
=
h_{K_{\perp}}
-
\lambda_{\perp}^{2},
\end{eqnarray}
where $h_{K_{z,\perp}}:=\tau_{c}K_{z,\perp}$ and
$h_{I}:=\tau_{c}I$.
We see that the Kondo couplings $\lambda_{z,\perp}$ 
renormalize the
exchange interactions between boundary states $K_{z,\perp}$,
making it ferromagnetic.
Then, the ground state of the boundary states
is polarized with the total spin equal to unity.
When we include $\Delta^{\dagger},\Delta$ in the RG analysis,
they also give rise to a contribution in the perturbative 
expansion.
In contrast to $\lambda_{z,\perp}$,
they suppress the ferromagnetic coupling between 
the boundary states $K_{z,\perp}$.
However,
the bulk interaction makes $\Delta^{\dagger},\Delta$ 
strongly irrelevant as stated above,
and hence they do not affect the RG flow.

\section{Conclusion}
\label{Conclusion}

To conclude,
we have investigated 
correlation effects of $(N,-N)$ CNT's with 
boundaries.
Low-energy effective Hamiltonians
were established 
by taking into account 
boundary states at the fermi energy.
Due to the special nature of the boundary states,
the differences between 
the Coulomb and the Hubbard interaction are found 
to be small.
We then discussed specific examples 
where only one or two boundary states appear.
In the infrared limit in RG analyses,
Kondo-like couplings between conduction electrons and boundary states
are shown to be vanishing,
and the bulk conduction electrons 
and boundary states are completely decoupled.
Doubly occupying a boundary state
becomes unfavorable in the infrared
near half filling
since the strong repulsive bulk interactions
renormalizes the interactions at the edge.
As the boundary states do not dissipate through
the coupling with conduction electrons,
they can be directly observed by local probes
such as scanning tunneling microscope.
The boundary states also manifest themselves in transport experiments
as
the conduction electrons and the boundary states 
give rise to independent contributions to, say,
tunneling density of states.

Furthermore,
the ground state at the boundary
is a highest spin state $S=1/2$ 
for the case of one boundary state
and $S=1$ for two boundary states,
since ferromagnetic couplings between the boundary states
are enhanced by the interactions between 
conduction electrons and the boundary states.
Then, the boundary states, as a whole, behave as a localized moment
which gives rise to a Curie-Weiss-like contribution
to the magnetic susceptibility,
apart from the bulk conduction electrons.

The result obtained here is consistent with 
spin polarization found in 
DMRG study
for a thin semi-conducting CNT
\cite{hikihara02},
and 
mean-field theory\cite{fujita96} or 
a LSDA calculation\cite{okada01}
for 2D sheet geometry.

At lower temperature,
the formation of a spin-gapped ground state
is suggested by the previous studies.
The fate of the spin polarization,
found here for the temperature above the gaps,
is left as an open question.
%Also, for a thicker CNT,
%situation is more complicated
%as there appear much more terms
%such as $H^{\hbox{\tiny{edge1}}}_{\hbox{\tiny{int}}}$ or
%$H^{\hbox{\tiny{edge3}}}_{\hbox{\tiny{int}}}$ 
%in effective Hamiltonians. 

Finally, we comment on an extra gapless mode
other than the modes for $k_{y}a_{0}=\pm 2\pi/3$ treated in the 
present paper.
When one considers the hybridization between
$\sigma-\pi$ orbitals,
which the simple tight-binding approximation adopted here
does not correctly capture,
there can appear a gapless mode for a very thin CNT,
as suggested by band calculations\cite{blase94,li01}.
In a realistic situation,
effects of the extra gapless mode
should be taken into account,
which can be possible along the lines of the present discussions.

\section*{Acknowledgments}

We are grateful to 
A. Furusaki,
H. Yoshioka,
K. Kusakabe,
M. Tsuchiizu,
R. Tamura and 
Y. Morita 
for useful discussions.
This work is supported by
JSPS (S.R.),
and
KAWASAKI STEEL 21st Century Foundation.
The computation in this work has been partly done 
at the Supercomputing Center, ISSP, University of Tokyo.


\begin{thebibliography}{99}

\bibitem{hamada92}
N. Hamada, S. Sawada, and A. Oshiyama,
\ Phys. \ Rev. \ Lett. \textbf{68}, 1579 (1992).

\bibitem{saito92}
R. Saito, M. Fujita, G. Dresselhaus, and M. S. Dresselhaus,
\ Appl. \ Phys. \ Lett. \textbf{60}, 2204 (1992).

\bibitem{mintmire92}
J. W. Mintmire, B. I. Dunlap, and C. T. White,
\ Phys. \ Rev. \ Lett. \textbf{68}, 631 (1992).

\bibitem{fujita96}
M. Fujita, K. Wakabayashi, K. Nakada, and K. Kusakabe,
\ J. \ Phys. \ Soc. \ Jpn. \textbf{65}, 1920 (1996).

\bibitem{ryu02}
S. Ryu and Y. Hatsugai,
\ Phys. \ Rev. \ Lett. \textbf{89}, 077002 (2002).

\bibitem{wakabayashi99}
K. Wakabayashi, M. Fujita, H. Ajiki, and M. Sigrist,
\ Phys. \ Rev. \ B \textbf{59}, 8271 (1999).

\bibitem{wakabayashi00}
K. Wakabayashi and M. Sigrist,
\ Phys. \ Rev. \ Lett. \textbf{84}, 3390 (2000).

\bibitem{wakabayashi01}
K. Wakabayashi,
\ Phys. \ Rev. \ B \textbf{64}, 125428 (2001).

\bibitem{wakabayashi98}
K. Wakabayashi, M. Sigrist, and M. Fujita,
\ J. \ Phys. \ Soc. \ Jpn. \textbf{67}, 2089 (1998).

\bibitem{okada01}
S. Okada and A. Oshiyama,
\ Phys. \ Rev. \ Lett. \textbf{87}, 146803 (2001).

\bibitem{hikihara02}
T. Hikihara, X. Hu, H. H. Lin, and C. Y. Mou,
cond-mat/0303159.

\bibitem{nakada98}
K. Nakada, M. Igami and M. Fujita, 
\ J. \ Phys. \ Soc. \ Jpn. \textbf{67}, 2388 (1998).

\bibitem{fujita97}
M. Fujita, M. Igami and K. Nakada,
\ J. \ Phys. \ Soc. \ Jpn. \textbf{66}, 1894 (1997).

\bibitem{lee02}
Y. L. Lee, Y. W. Lee,
\ Phys. \ Rev. \ B \textbf{66}, 245402 (2002).

\bibitem{egger97}
R. Egger and A. O. Gogolin,
\ Phys. \ Rev. \ Lett. \textbf{79}, 5082 (1997).

\bibitem{egger98}
R. Egger and A. O. Gogolin,
\ Eur. \ Phys. \ J. \ B \textbf{3}, 281 (1998).

\bibitem{yoshioka99}
H. Yoshioka and A. A. Odintsov,
\ Phys. \ Rev. \ Lett. \textbf{82}, 374 (1999).

\bibitem{odintsov99}
A. A. Odintsov and H. Yoshioka,
\ Phys. \ Rev. \ B \textbf{59}, R10457 (1999).

\bibitem{krotov97}
Y. A. Krotov, D. H. Lee, and S. G. Louie,
\ Phys. \ Rev. \ Lett. \textbf{78}, 4245 (1997).

\bibitem{balents97}
L. Balents and M. P. A. Fisher,
\ Phys. \ Rev. \ B \textbf{55}, R11973 (1997).

\bibitem{lin98}
H. H. Lin,
\ Phys. \ Rev. \ B \textbf{58}, 4963 (1998).

\bibitem{kane97}
C. Kane, L. Balents, and M. P. A. Fisher,
\ Phys. \ Rev. \ Lett. \textbf{79}, 5086 (1997).

\bibitem{bockrath99}
M. Bockrath, D. H. Cobden, J. Lu, A. G. Rinzler,
R. E. Smalley, L. Balents, and P. L. McEuen,
\ Nature (London) \textbf{397}, 598 (1999).

%\bibitem{cobden98}
%D. H. Cobden, M. Bockrath, P. L. McEuen,
%A. G. Rinzler, and R. E. Smalley,
%\ Phys. \ Rev. \ Lett. \textbf{81}, 681 (1998).

%\bibitem{tans98}
%S. J. Tans, M. H. Devoret, R. J. A. Groeneveld,
%and C. Dekker,
%\ Nature \textbf{394}, 761 (1998).

\bibitem{kane92}
C. L. Kane and M. P. A. Fisher,
\ Phys. \ Rev. \ Lett. \textbf{68}, 1220 (1992);
\ Phys. \ Rev. \textbf{46}, 15233 (1992).

\bibitem{egger92}
S. Eggert and I. Affleck,
\ Phys. \ Rev. \ B. \textbf{46}, 10866 (1992).

\bibitem{furusaki93}
A. Furusaki and N. Nagaosa,
\ Phys. \ Rev. \ B. \textbf{47}, 4631 (1993).

\bibitem{wang94}
E. Wang and I. Affleck,
\ Nucl. \ Phys. \ B. \textbf{417}, 403 (1994).

\bibitem{fabrizio95}
M. Fabrizio, A. O. Gogolin,
\ Phys. \ Rev. \ B \textbf{51}, 17827 (1995).

\bibitem{yoshioka02}
H. Yoshioka and Y. Okamura,
\ J. \ Phys. \ Soc. \ Jpn. \textbf{71}, 2512 (2002).

\bibitem{yoshioka02-2}
H. Yoshioka,
LT23 Proceedings.

\bibitem{kusakabe02}
K. Kusakabe and M. Maruyama,
\ Phys. \ Rev. \ B. \textbf{67}, 092406 (2003).

\bibitem{si93}
Q. Si and G. Kotliar,
\ Phys. \ Rev. \ B. \textbf{48}, 13881 (1993).

\bibitem{schotte69}
K. D. Schotte and U. Schotte,
\ Phys. \ Rev. \textbf{182}, 479 (1969).

\bibitem{rem1}
Although superconducting pairing operators
$\Delta^{\dagger}$, $\Delta$ do not affect the 
global features of the RG flows,
they can renormalize couplings $h_{\epsilon}$ and $h_{U}$
at $l=0$,
which might result in a different conclusion,
e.g.,
$U_{e}+2\epsilon_{e}\to -\infty$.
We derived RG equations including the effects of 
$\Delta^{\dagger}$ and $\Delta$ to check that
this does not occur.

\bibitem{blase94}
X. Blase, L. X. Benedict, E. L. Shirley, and S. G. Louie,
\ Phys. \ Rev. \ Lett. \textbf{72}, 1878 (1994).

\bibitem{li01}
Z. M. Li, Z. K. Tang, H. J. Liu, N. Wang, C. T. Chan,
R. Saito, S. Okada, G. D. Li, J. S. Chen,
N. Nagasawa, and S. Tsuda,
\ Phys. \ Rev. \ Lett. \textbf{87}, 127401 (2001).


\end{thebibliography}
\end{document}